\documentclass[%
 pdftex,
 10pt,
 letterpaper,
 prd,
 longbibliography,
 floatfix,
 twocolumn,
 superscriptaddress,
 nofootinbib,
 amsfonts,
 amssymb,
 amsmath
]{revtex4-2}

\usepackage{graphicx}
\usepackage{cmap} 
\usepackage{ifpdf}
 \ifpdf
 \usepackage{cmap} 
 \usepackage{color,graphicx}
 \usepackage{epstopdf} 
 \else
 \usepackage{color,graphicx}
 \fi
\usepackage{bm}
\usepackage[dvipsnames]{xcolor}
\usepackage{hyperref}
\hypersetup{%
colorlinks=true,%
citecolor=Blue,%
filecolor=Blue,
linkcolor=Blue,
urlcolor=Magenta 
}
\usepackage[activate={true,nocompatibility},final,tracking=true,kerning=true,spacing=true,factor=1100,stretch=10,shrink=10]{microtype}

\graphicspath{{./}}

\newcommand{\mara}{{MARATHON}}

\newcommand{\ud}  {\mathrm{d}}
\newcommand{\gev} {\ \mathrm{GeV}}

\newcommand{\gevsq}{\ \mathrm{GeV}^2}

\newcommand{\ceps}{\varepsilon}

\newcommand{\eq}[1]{Eq.~(\ref{#1})}
\newcommand{\Eqs}[2]{Eq.~(\ref{#1}) and (\ref{#2})}

\newcommand{\htri}{{}^3\text{H}}
\newcommand{\hetri}{{}^3\text{He}}

\allowdisplaybreaks

\begin{document}

\title{Off-shell modifications of bound nucleons and parton distributions}

\author{S.~I.~Alekhin}
\affiliation{II. Institut f\"ur Theoretische Physik, Universit\"at Hamburg,
D--22761 Hamburg, Germany}
\author{S.~A.~Kulagin}
\affiliation{Institute for Nuclear Research of the Russian Academy of Sciences,
117312 Moscow, Russia}
\author{R.~Petti}
\affiliation{Department of Physics and Astronomy,
University of South Carolina, Columbia, South Carolina 29208, USA}


\begin{abstract}
\noindent
We discuss results from our global QCD analyses including nuclear data
off deuterium from various measurements, as well as off  $\htri$ and $\hetri$ targets from the \mara{} experiment. We simultaneously determine the parton
distribution functions of the proton, the higher-twist terms, and the nucleon off-shell correction functions responsible
for the modifications of the partonic structure in bound protons and neutrons. 
In particular, we study the neutron-proton asymmetry of the off-shell correction and its interplay with the treatment of the higher-twist terms.
We observe that the data on the $\hetri/\htri$ cross section ratio are consistent with a single isoscalar off-shell function.
We also provide our predictions on the ratio $F_2^n/F_2^p$ and on the $d$ and $u$ quark distributions in the proton and in the $\htri$ and $\hetri$ nuclei.
\end{abstract}

\maketitle

\section{Introduction}
\renewcommand{\thefootnote}{\fnsymbol{footnote}}

Using data from 
deep-inelastic scattering (DIS) off nuclear targets with different proton-neutron content in global QCD analyses allows to unravel the physics mechanisms responsible of the modifications of bound nucleons in the nuclear environment,
to accurately constrain PDFs in the neutron, as well as to test the nucleon charge symmetry. 
We summarize the results of our recent global QCD analyses~\cite{Alekhin:2022tip,Alekhin:2022uwc},
in which we simultaneously constrain the proton PDFs, the higher-twist (HT) terms, and the functions describing the modification of the nucleon structure functions (SFs) in nuclei.%
\footnote{Presented at DIS2023: XXX International Workshop on Deep-Inelastic Scattering and Related Subjects, Michigan State University, USA, 27-31 March 2023.} 
We use deuterium DIS data from various experiments,
the data on the $\hetri/\htri$ cross section ratio from the \mara{} experiment~\cite{MARATHON:2021vqu},
along with a typical set of the proton DIS and collider data (for details see Refs.~\cite{Alekhin:2017kpj,Alekhin:2022tip}).
Nuclear corrections are treated following the microscopic model of Ref.~\cite{Kulagin:2004ie}, which addresses a number of effects relevant in different kinematical regions of Bjorken $x$.
In the large-$x$ region relevant for the nuclear DIS data considered,
the most important nuclear corrections originate from the nuclear momentum distribution, the nuclear binding~\cite{Akulinichev:1985ij,Kulagin:1989mu} and the off-shell (OS) corrections to the bound nucleon SFs~\cite{Kulagin:1994fz,Kulagin:2004ie}.
The latter are directly related to the modification of the partonic structure of bound nucleons, and the validity of such approach 
was demonstrated in the analysis of data on the nuclear EMC effect~\cite{Kulagin:2004ie}.
The observations of Ref.~\cite{Kulagin:2004ie} have been confirmed in a global QCD analysis including deuterium DIS data~\cite{Alekhin:2017fpf,Alekhin:2022tip}.

The data from the \mara{} experiment on DIS cross sections off $\htri$ and $\hetri$ targets allow to constrain the nucleon isospin dependence of the OS functions~\cite{Alekhin:2022uwc}.
The OS functions, in turn, determine the in-medium modifications of the partonic structure of bound protons and neutrons. 
It should be noted, that most of the fixed-target nuclear data in the present analysis
typically have invariant momentum transfer squared $Q^2$ about a few $\gevsq$ and for this reason the HT terms should be addressed.
To this end, we consider two different models of HT terms and study the interplay between the underlying HT model and the resulting predictions on the ratio $d/u$ of the quark distributions, the structure function ratio $F_2^n/F_2^p$, and the proton-neutron asymmetry in the off-shell correction.

\section{Theory background}

The cross sections of the spin-independent charged-lepton inelastic scattering are fully described in terms of $F_T=2xF_1$ and $F_2$ SFs.
In the DIS region of high invariant momentum transfer squared $Q^2$,
SFs can be expressed as a power series in $Q^{-2}$ (twist expansion) within the operator product expansion (OPE).
The leading twist (LT) SFs are given by a convolution of PDFs with the functions describing the quark-gluon interaction at the scale $Q$, which can be computed perturbatively as a series in the strong coupling constant (see, e.g.,~\cite{Accardi:2016ndt}).
SFs can then be writen as
\begin{equation}\label{eq:sf}
F_i = F_i^{\text{TMC}} + H_i/ Q^2 + \cdots,  
\end{equation}
where $i=T,2$, $F_i^{\text{TMC}}$ are the corresponding LT SFs including the target mass correction (TMC)~\cite{Georgi:1976ve},
$H_i$ describes the twist-4 contribution.
We consider two HT models commonly used:
(i) additive HT model (aHT) motivated by the OPE, in which $H_i=H_i(x)$  and
(ii) multiplicative HT model (mHT)~\cite{Virchaux:1991jc},
in which $H_i$ is assumed to be proportional to the corresponding LT SF, $H_i=F_i^\text{LT}(x,Q^2) h_i(x)$.

We address nuclear corrections in the DIS process by treating it as an incoherent scattering 
off bound nucleons in the target rest frame. 
The deuteron SFs can be calculated as the sum of bound proton and neutron SFs convoluted with the  nucleon momentum distribution given by the deuteron wave function squared,
$\left|\Psi_d(\bm k)\right|^2$:
\begin{align}
	\label{eq:IA2}
	F_i^d = \int \ud^3\bm k K_{ij} \left|\Psi_d(\bm k)\right|^2 \left(F_j^p + F_j^n\right),
\end{align}
where the integration is performed over the bound nucleon momentum $\bm k$, $i,j=T,2$, we assume a summation over the repeated index $j$, and $K_{ij}$ are the kinematic factors~\cite{Kulagin:2004ie,Alekhin:2022tip}.
For nuclei with $A\ge 3$
the convolution by \eq{eq:IA2} requires the integration over the energy spectrum of the residual nuclear system, along with the nucleon momentum, which are described by the nuclear spectral functions $\mathcal P_{p/A}$ and $\mathcal P_{n/A}$~\cite{Akulinichev:1985ij,Kulagin:1989mu,Kulagin:1994fz,Kulagin:2004ie,Kulagin:2010gd}:
\begin{align}
\label{eq:IA}
F_i^A = \int\!\ud^4 k K_{ij} \left(\mathcal{P}_{p/A} F_j^p + \mathcal{P}_{n/A} F_j^n\right),
\end{align}
where the integration is performed over the bound nucleon four-momentum $k$.
The corresponding nucleon off-shell SFs in both \eq{eq:IA2} and \eq{eq:IA} depend on the scaling variable $x'=Q^2/2k\cdot q$,
the DIS scale $Q^2$, and the nucleon invariant mass squared $k^2=k_0^2-\bm k^2\not=M^2$, where $M$ is the nucleon mass.
This latter dependence originates from both the power TMC terms of the order $k^2/Q^2$ and the OS dependence of the LT SFs.
Following Refs.~\cite{Kulagin:1994fz,Kulagin:2004ie},
we treat the OS correction in the vicinity of the mass shell $k^2=M^2$ by expanding SFs in a power series in $v=(k^2-M^2)/M^2$.
To the leading order in $v$ we have
\begin{align}
\label{SF:OS}
F_i^\text{LT}(x,Q^2,k^2) &= F_i^\text{LT}(x,Q^2,M^2)\left( 1+\delta f_i\,v \right),
\\
\label{eq:deltaf}
\delta f_i &= \partial \ln F_i^\text{LT}(x,Q^2,k^2)/\partial \ln k^2,
\end{align}
where the derivative is taken on the mass shell $k^2=M^2$.
We assume equal functions $\delta f_T=\delta f_2=\delta f$ for $F_T$ and $F_2$,
motivated by the observation that $F_T\approx F_2$ in the region for which the OS effect is numerically important~\cite{Kulagin:2004ie,Kulagin:2010gd,Alekhin:2017fpf,Alekhin:2022tip}.

We use \Eqs{eq:IA2}{eq:IA} to address the nuclear corrections from the momentum distribution, the nuclear binding, and the OS effect, which are the main nuclear corrections at large $x$.
Other nuclear effects like the meson-exchange currents and the nuclear shadowing 
result in corrections comparable to the 
experimental uncertainties at large $x$~\cite{Alekhin:2017fpf} and are therefore 
neglected in the present analysis. 
We use a deuteron wave function based on the Argonne nucleon-nucleon otential~\cite{Wiringa:1994wb,Veerasamy:2011ak} (AV18).
For the $A=3$ nuclei, the proton (neutron) spectral function $\mathcal P_{p(n)/A}(\ceps,\bm k)$ describes the corresponding energy ($\ceps=k_0-M$) and momentum ($\bm k$) distribution in a nucleus at rest.
The nuclear spectral function involves contributions from all possible $A-1$ intermediate states.
For the proton spectral function of $^3$He, $\mathcal P_{p/\hetri}$,
the relevant contributions come from two-body $pn$ intermediate states,
both the $pn$ continuum and the $pn$ bound state, i.e. the deuteron.
The neutron spectral function of $^3$He, $\mathcal P_{n/\hetri}$, involves only the $pp$ continuum states.
Similarly, for the $^3$H nucleus, the neutron spectral function involves contributions
from the bound $pn$ state and from the $pn$ continuum states, while
the proton spectral function includes only the $nn$ continuum states.
We use the $^3$He and $^3$H spectral functions of Ref.~\cite{Pace:2001cm}
computed with the AV18 nucleon-nucleon force and accounting for the Urbana three-nucleon interaction, as well as the Coulomb effect in $\hetri$.
The details of the corresponding 
nuclear convolution equation, \Eqs{eq:IA2}{eq:IA}, can be found in Refs.~\cite{Kulagin:2004ie,Kulagin:2010gd,Alekhin:2022tip,Alekhin:2022uwc}.

\section{Analysis framework}

We simultaneously constrain the proton PDFs, the HT corrections, and the proton and the neutron OS functions, $\delta f^p$ and $\delta f^n$, describing the modifications the proton and neutron PDFs in the nuclear environment,
in a global QCD analysis. 
The datasets used are described in Refs.~\cite{Alekhin:2017fpf,Alekhin:2022tip} 
and include charged-lepton DIS data off proton, deuterium, $\htri$, and $\hetri$ targets, as well as data from the $W^\pm/Z$ boson production at hadron colliders.
In particular, data on the ratio of the DIS cross sections of the three-body nuclei, $\sigma^{\hetri}/\sigma^{\htri}$, from the \mara{} experiment~\cite{MARATHON:2021vqu} allow to study
the neutron-proton asymmetry  $\delta f^a=\delta f^n-\delta f^p$~\cite{Alekhin:2022uwc}.

We parametrize the proton PDFs following Ref.~\cite{Alekhin:2017fpf}, while the $Q^2$ dependence of the LT SFs is computed at the next-to-next-to-leading order (NNLO) in perturbative QCD.
The functions $H_i(x)$ in the aHT model are treated independently for $i=T,2$ and are parameterized in the form of spline polynomials.
A similar procedure is applied for the functions $h_i$ in the mHT model.
To reduce the number of parameters we assume
$H_i^p=H_i^n$ in the aHT model and also test the assumption $h_i^p=h_i^n$ in the mHT model.
We apply the cuts  $Q^2>2.5$ and $W>1.8\gev$, where $W$ is the invariant mass of the produced hadronic states. Additional details about the analysis setup, like the treatment of the uncertainties and the PDFs and HTs parametrizations, can be found in Refs.~\cite{Alekhin:2022tip,Alekhin:2022uwc}. 

We parametrize the proton function $\delta f^p(x)$ in terms of a generic second order polynomial~\cite{Alekhin:2017fpf,Alekhin:2022tip}:
\begin{equation}\label{eq:dfpar1}
\delta f^p(x)=a+bx+cx^2,
\end{equation}
where the parameters $a$, $b$, and $c$ are determined simultaneously with those of the proton PDFs and HTs.
We also consider the corresponding neutron-proton asymmetry $\delta f^a=\delta f^n-\delta f^p$, for which we assume a linear function, $\delta f^a(x)=a_1+b_1x$, 
with $a_1$ and $b_1$ free parameters.

\section{Results and discussion}

In order to study the impact of various effects, we perform a number of fits with different settings.
In our default QCD analysis we assume equal off-shell functions for protons and neutrons, $\delta f^p=\delta f^n=\delta f$, and the aHT model for the HT terms. With such settings we obtain~\cite{Alekhin:2022uwc} a good agreement with the \mara{} data on the ratio $\sigma^{\hetri}/\sigma^{\htri}$~\cite{MARATHON:2021vqu}
with $\chi^2$ per number of data points (NDP) of $20/22$, and $\chi^2/\text{NDP}=4861/4065$ considering all data~\cite{Alekhin:2022tip,Alekhin:2022uwc}.

\begin{figure*}[hbt]
\centering
\vspace{-2ex}
\hspace{-1em}%
\includegraphics[width=0.47\linewidth]{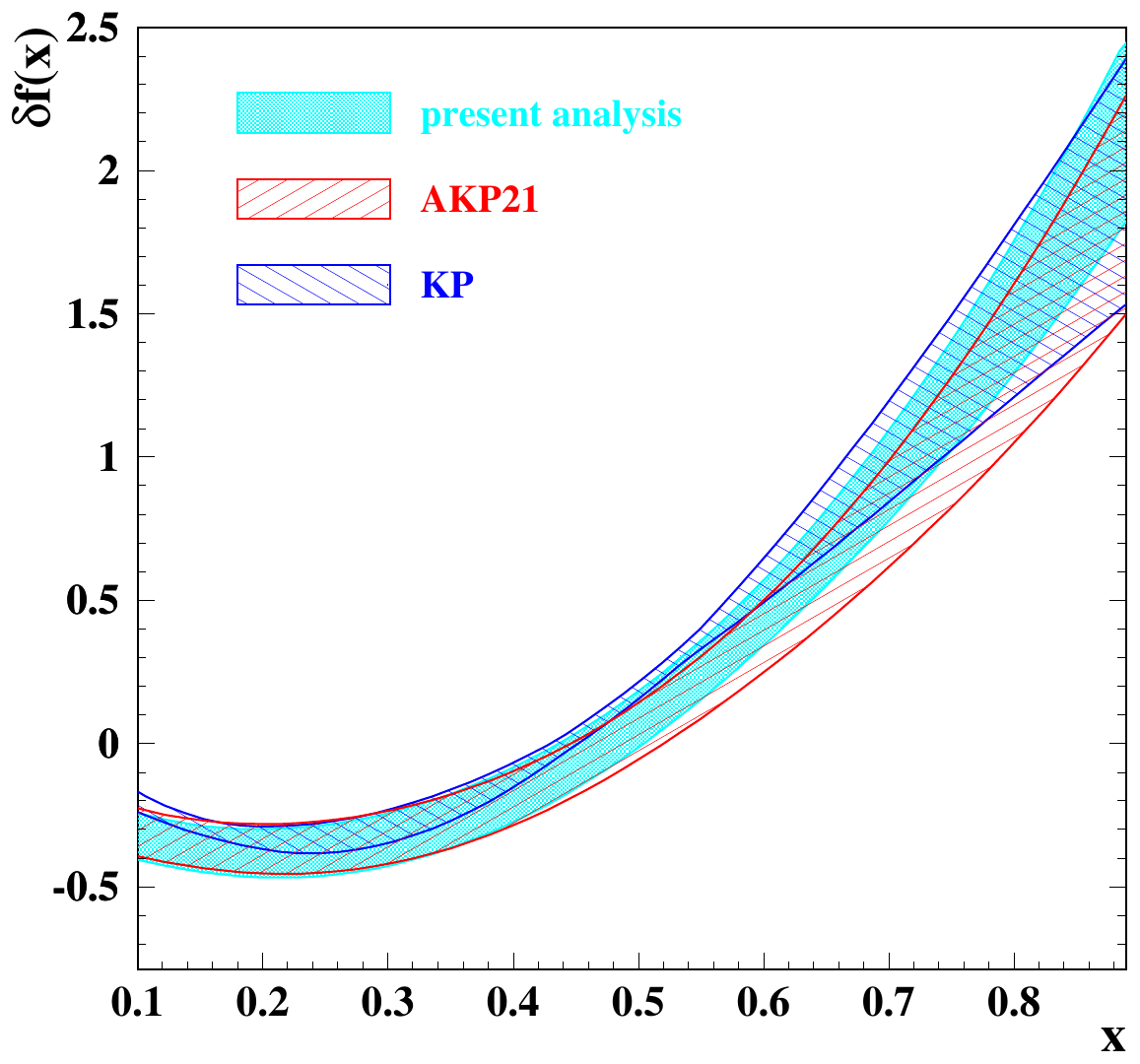}%
\hspace{1em}%
\includegraphics[width=0.47\linewidth]{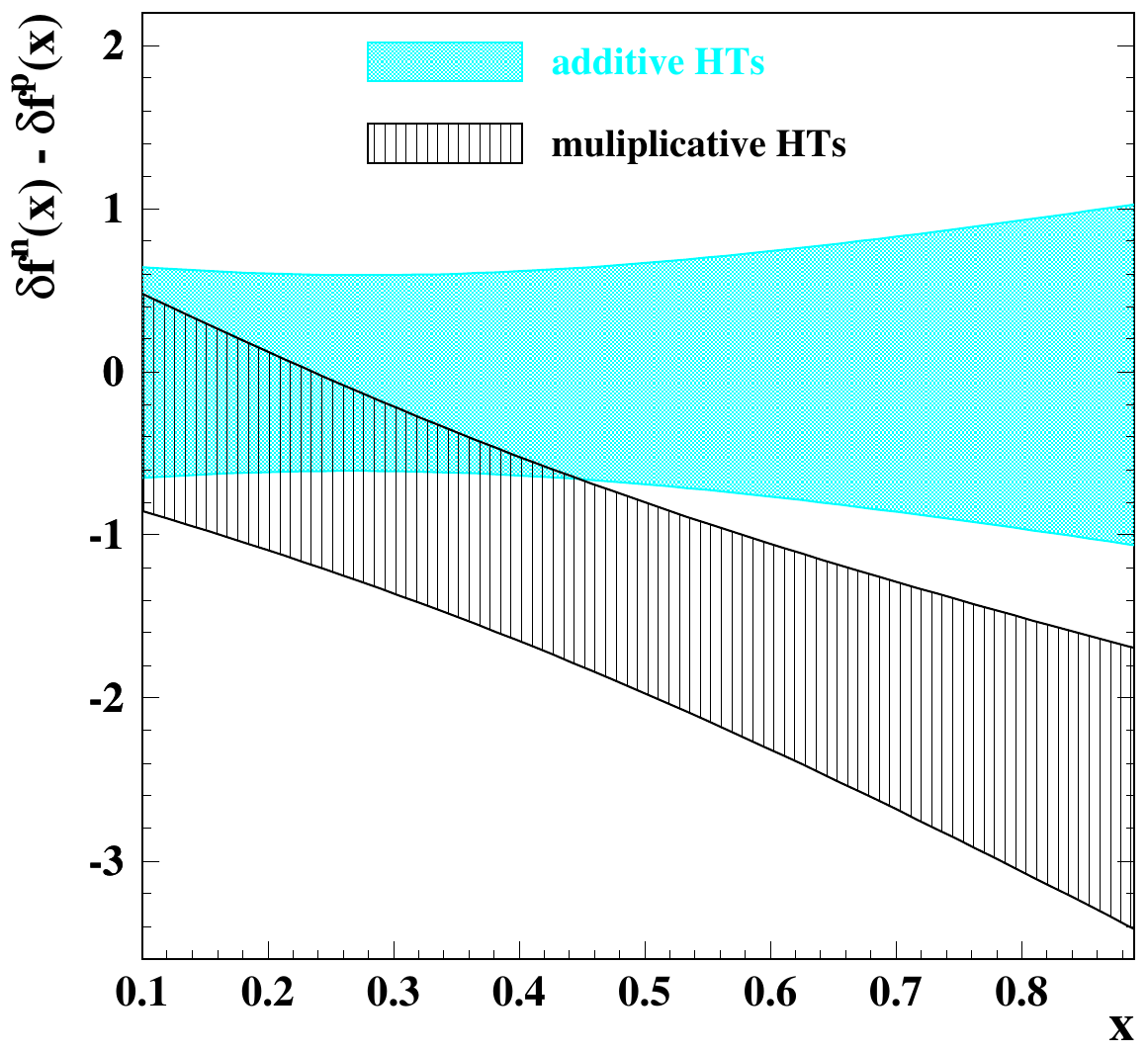}
\caption{Left:
The $1\sigma$  uncertainty band on the OS function obtained assuming $\delta f^p=\delta f^n$ and the aHT model for the HT terms (shaded area)~\cite{Alekhin:2022uwc}.
Also shown are the results of Refs.~\cite{Kulagin:2004ie} (KP) and \cite{Alekhin:2022tip} (AKP21).
Right:
The $1\sigma$ uncertainty band on the neutron-proton asymmetry $\delta f^n(x)-\delta f^p(x)$ for the aHT (shaded area) and mHT (hashed area) models~\cite{Alekhin:2022uwc}.}
\label{fig:deltaf}
\end{figure*}

The function $\delta f(x)$ obtained from the analysis of Ref.~\cite{Alekhin:2022uwc} is shown in Fig.~\ref{fig:deltaf} (left panel).
The results are in good agreement with the original determination~\cite{Kulagin:2004ie} from the ratios $\sigma^A/\sigma^d$ of the DIS cross sections
off nuclear targets with a mass number $A\geq 4$ using the proton and the neutron SFs of Ref.~\cite{Alekhin:2007fh}.
The results of Ref.~\cite{Alekhin:2022uwc} also agree with those of Ref.~\cite{Alekhin:2022tip},
which does not include the \mara{} data from $A=3$ nuclei.
It should be also noted that the data on the ratio $\sigma^{\hetri}/\sigma^{\htri}$
allows a reduction of the $\delta f(x)$ uncertainty at large $x$.

The results shown in Fig.~\ref{fig:deltaf} (left panel) are obtained assuming an isospin-symmetric function $\delta f^p=\delta f^n$ and the aHT model for the HT terms.
Such an assumption was verified in the analysis of the EMC effect in Ref.~\cite{Kulagin:2004ie} and was also used in Refs.~\cite{Alekhin:2017fpf,Alekhin:2022tip}.
The \mara{} data on the ratio $\sigma^{\hetri}/\sigma^{\htri}$ were used to constrain the asymmetry $\delta f^a=\delta f^n-\delta f^p$~\cite{Alekhin:2022uwc}.
With the aHT model we obtain a function 
$\delta f^p$ similar to that of the isospin-symmetric case shown in Fig.~\ref{fig:deltaf} (left panel), as well as an asymmetry $\delta f^a$ consistent with zero within uncertainties, as shown in Fig.~\ref{fig:deltaf} (right panel)~\cite{Alekhin:2022uwc}.
However, we obtain substantially 
different results on the function $\delta f^a$ with the mHT model, as shown in the right panel of Fig.~\ref{fig:deltaf}.

\begin{figure}[htb]
\centering
\hspace{-1em}%
\includegraphics[width=1.0\linewidth]{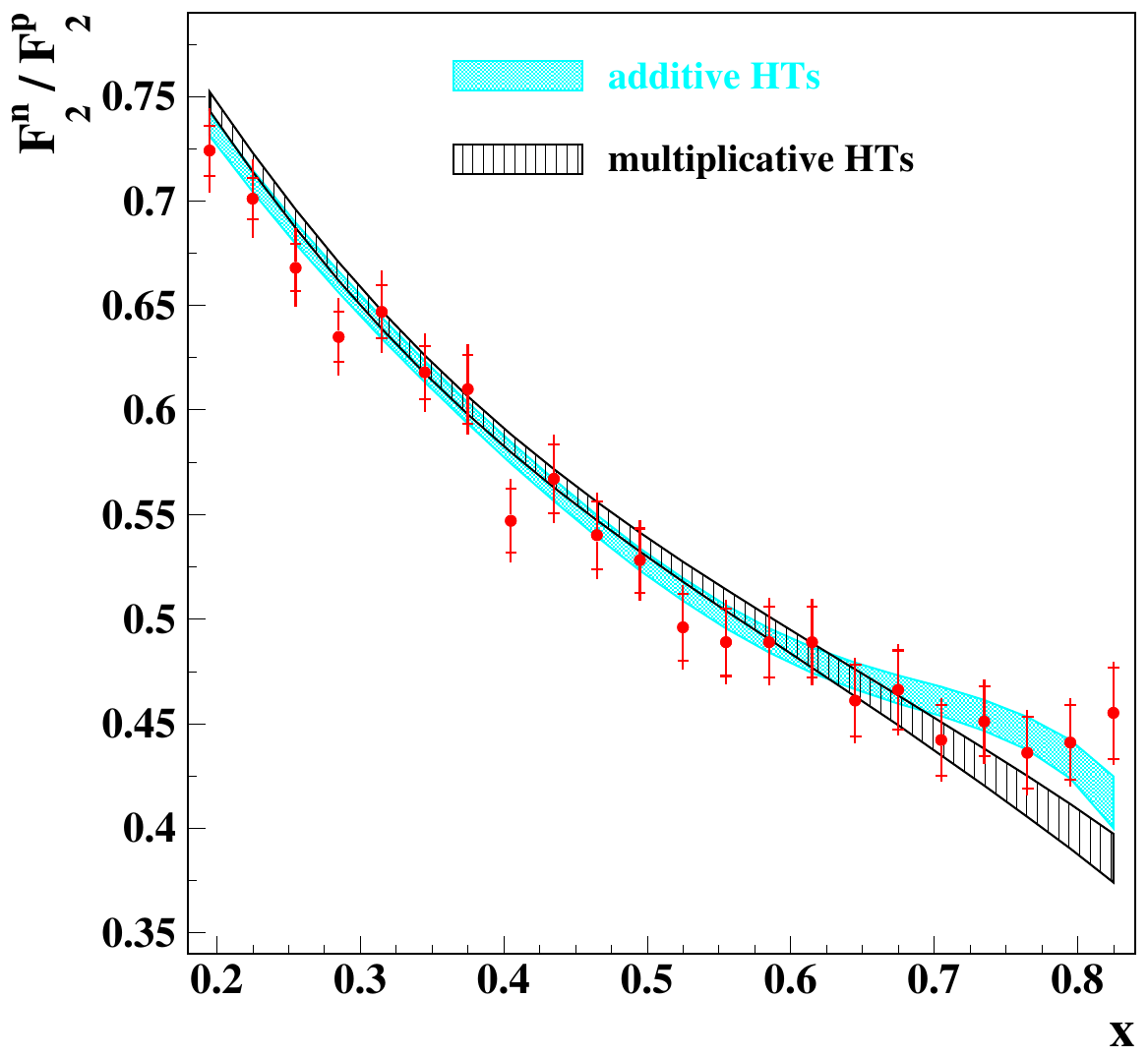}
\caption{%
The \mara{} data on $F_2^n/F_2^p$~\cite{MARATHON:2021vqu} vs. the results of Ref.~\cite{Alekhin:2022uwc}
($1\sigma$ uncertainty band) with the aHT model (shaded area) and the mHT model (hashed area).}
\label{fig:f2nf2p}
\vspace{-1ex}
\end{figure}

The underlying reason for a nonzero asymmetry $\delta f^a$ in the mHT model is 
the interplay between the HT terms and the LT ones, as $H_i=F_i^\text{LT}(x,Q^2)h_i(x)$. 
On one side, the factor $F_i^\text{LT}$ results in a $Q^2$ dependence in $H_i$, as illustrated in Ref.~\cite{Alekhin:2022tip}. 
On another side, the factor $F_i^\text{LT}$ also introduces an explicit isospin dependence in the HT terms, present even with an isoscalar function $h_i^p=h_i^n$. The nonzero asymmetry $\delta f^a$ we found in the mHT model (Fig.~\ref{fig:deltaf}) may therefore be a bias partially compensating such an isospin dependence. 

\begin{figure}[htb]
\vspace{-4ex}
\centering
\includegraphics[width=1.0\linewidth]{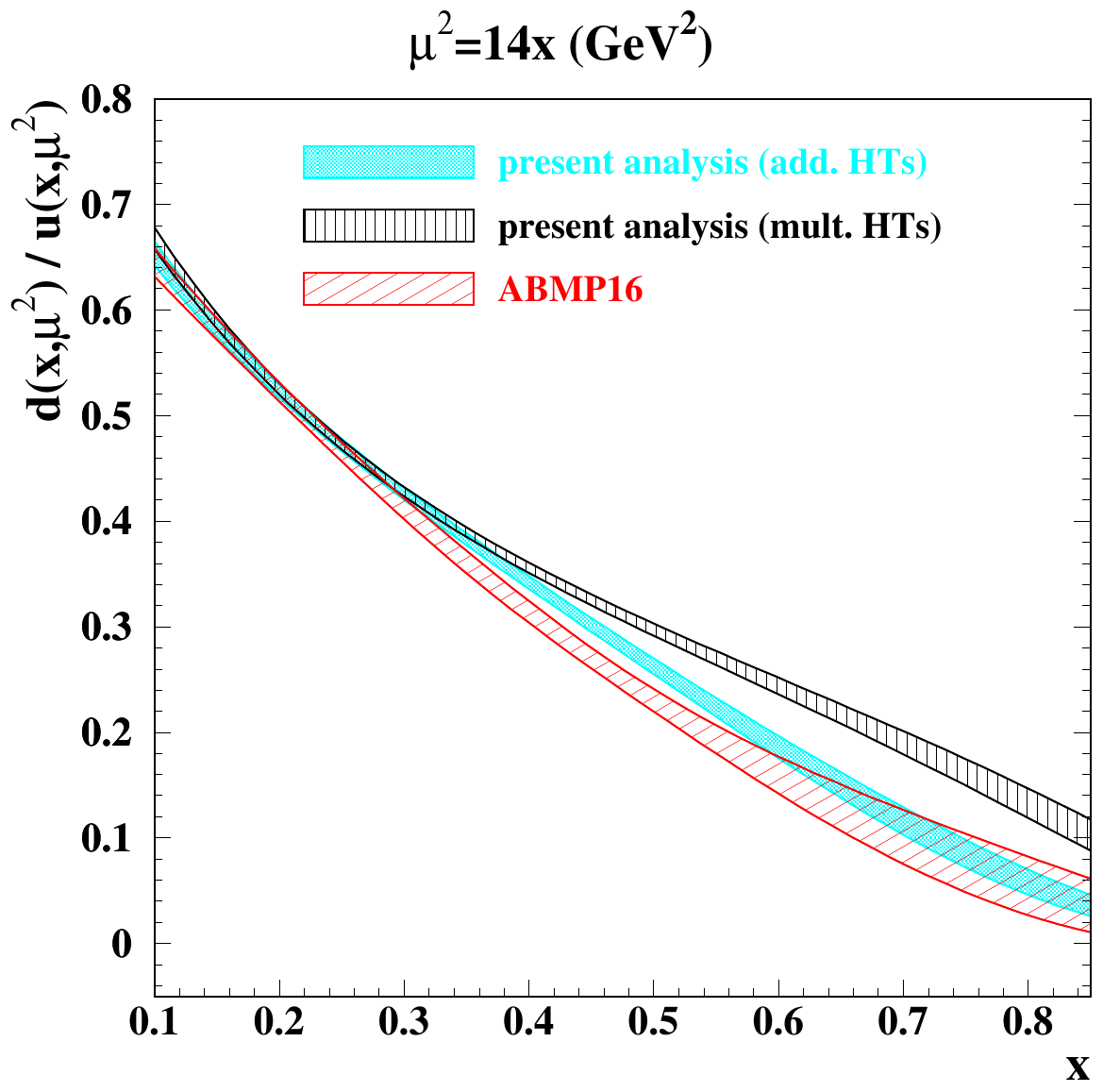}
\caption{%
Comparison of the $1\sigma$ uncertainty bands for the $d/u$ ratio in the proton  obtained with the aHT (shaded area) and mHT (right-tilted hashed area) models at the factorization scale $\mu^2=14x\ (\text{GeV}^2)$ in AKP22 analysis~\cite{Alekhin:2022uwc}.
Also shown are the results of ABMP16 analysis~\cite{Alekhin:2017kpj} (left-tilted hashed area), which does not include any nuclear data.
}\label{fig:du}
\end{figure}

The \mara{} data are particularly interesting as they are sensitive not only to isospin effects, but also to the HT contributions in the region $x >0.6$. 
Figure~\ref{fig:f2nf2p} shows a comparison of the \mara{} $F_2^n/F_2^p$ measurement with our predictions for both the aHT and the mHT models. 
Overall, we obtain an excellent description of the \mara{} data using our default QCD analysis with the aHT model, with a $\chi^2/\text{NDP}=20/22$. 
The data seem to prefer the aHT model over the mHT, as indicated by the higher value of  $\chi^2/\text{NDP}=34/22$ with the latter. 

We compare the ratio of $d/u$ quark distributions obtained with different HT models for the kinematics of the \mara{} experiment, 
together with the one from the analysis of Ref.~\cite{Alekhin:2017kpj} (ABMP16), which was performed with the aHT model but without any nuclear data (see Fig.~\ref{fig:du}).
In the latter case the ratio $d/u$ is mostly constrained by forward $W$-boson production data from the LHCb~\cite{LHCb:2015kwa,LHCb:2015okr,LHCb:2015mad} and D0~\cite{D0:2014kma} experiments. The ABMP16 result is in good agreement with the present one obtained with the aHT model.
Instead, the $d/u$ ratio in the mHT model is substantially higher at large $x$. Such an enhancement appears to be correlated with the nonzero values of the 
asymmetry $\delta f^a$ (cf. Figs.~\ref{fig:deltaf} and \ref{fig:du}). This observation indicates a tension between the DIS and Drell-Yan data in the mHT model.

We can use the results of Ref.~\cite{Alekhin:2022uwc} to calculate the nuclear modifications of 
the quark distributions for different flavors.
In particular, the nuclear PDFs $q_{i/A}$
for the parton type $i=u,d,\ldots$ can be obtained from the proton and neutron PDFs using a convolution equation similar to \eq{eq:IA}~\cite{Kulagin:2014vsa,Kulagin:2016fzf}:
\begin{align}
xq_{i/A} &=\! \int\! \ud^4k\! \left(1+\frac{k_z}{M}\right)\!
\left(\mathcal P_{p/A} x'q_{i/p} + \mathcal P_{n/A} x'q_{i/n}\right),
\label{eq:convq}
\end{align}
where the off-shell nucleon PDFs depend on $x'$, $Q^2$, and $k^2$, and the $z$-axis is antiparallel to the momentum transfer $\bm q$. The corresponding 
off-shell corrections are treated as in \Eqs{SF:OS}{eq:deltaf} with an OS function $\delta f_q$ which, in general, depends on the quark flavor. 
We use the results we obtained with the aHT default model, suggesting the same 
OS function $\delta f_q=\delta f$ for both $u$ and $d$ quark distributions. 

\begin{figure*}[hbt]
\centering
\hspace{-1em}%
\includegraphics[width=0.5\linewidth]{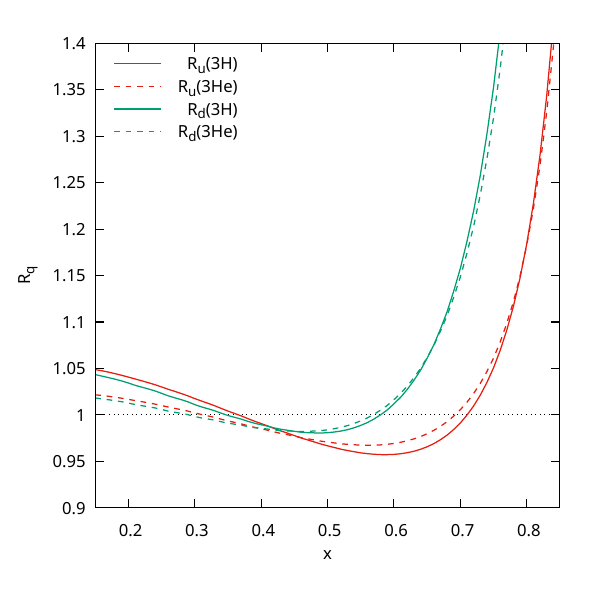}%
\hspace{1em}%
\includegraphics[width=0.5\linewidth]{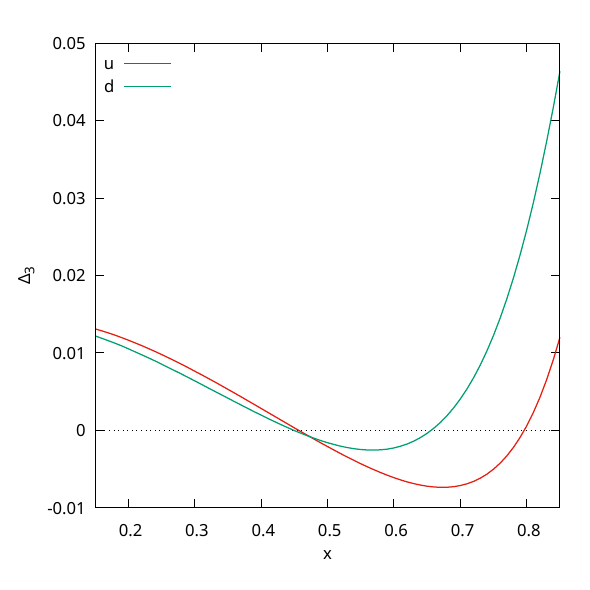}
\caption{Left:
The ratio $R_u$ and $R_d$ computed using the results of analysis~\cite{Alekhin:2022uwc} at $Q^2=10\gevsq$ for $\htri$ and $\hetri$ nuclei.
Right:
The $\htri-\hetri$ asymmetry $\Delta_3$ for $u$ and $d$ valence quark distributions in bound proton.
}
\label{fig:ud3}
\end{figure*}

We calculated the ratio $R_q=q_{p/A}/q_p$ 
between the proton contribution $q_{p/A}$ to \eq{eq:convq} and the corresponding free proton PDF for both $u$ and $d$ quarks in $\hetri$ and $\htri$ using the proton PDFs and the $\delta f(x)$ function of Ref.~\cite{Alekhin:2022uwc}, shown in Fig.~\ref{fig:ud3}.
The ratio $R_q$ describes the modifications of the parton distributions $q=u,d,\ldots$ in a bound proton due to the energy-momentum distribution  and to the off-shell effect.
Even using an isoscalar OS function $\delta f$, 
we observe a pronounced flavor dependence of the EMC effect at $x>0.5$ as a result of the 
convolution of PDFs with different $x$ dependence with the nucleon momentum distribution.
The nuclear dependence of $R_q$ is also noticeable and is owed to the differences in the proton spectral functions of $\htri$ and $\hetri$. 
In order to further clarify the flavor dependence of nuclear effects, we show the asymmetry $\Delta_3=(R_q(\htri)-R_q(\hetri))/(R_q(\htri)+R_q(\hetri))$ for both $q=u$ and $d$ quarks in the right panel of Fig.~\ref{fig:ud3}.

The results described above contrast with those of Ref.~\cite{Cocuzza:2021rfn}, 
claiming a significant isovector nuclear EMC effect from a global QCD analysis including $A=2$ and $A=3$ DIS data (see Fig.~3 in \cite{Cocuzza:2021rfn}).
We comment in this context that Ref.~\cite{Cocuzza:2021rfn} uses the mHT model of HT terms in their analysis.
As we show above (see Ref.~\cite{Alekhin:2022uwc}), there is an interplay of the nucleon isospin dependence of the OS correction and the $d/u$ ratio and the HT terms in the mHT model. In particular, the isospin effect in the OS correction tends to compensate the isospin dependence of the HT terms in the mHT model.
Furthermore, the analysis of Ref.~\cite{Cocuzza:2021rfn} 
introduces an explicit nuclear dependence in the 
OS functions for individual quark flavors, which 
may result in additional correlations among parameters potentially affecting the results.

We note that the HT terms cancel out in the ratio $F_i^n/F_i^p = F_i^{LT, n}/F_i^{LT, p}$ in the mHT model with the assumption $h_i^p=h_i^n$.
We therefore expect that analyses of the \mara{} $^3$He/$^3$H ratio based on a naive LT approximation for SFs~\cite{Segarra:2021exb} 
could be affected by somewhat similar biases on the 
resulting isospin dependence of the OS function $\delta f$ as the ones found in the mHT model.

\section{Summary and outlook}

We obtain a good description of the \mara{} data within the simple assumption of isoscalar HT contributions in the aHT model. 
From our global QCD analysis
we get the same  function $\delta f$ for both the protons and the neutrons, within uncertainties.
This result is consistent with our former observations from the global QCD analyses including deuterium DIS data~\cite{Alekhin:2017fpf,Alekhin:2022tip}, as well as with the analysis of the nuclear DIS data with
$A\geq 3$~\cite{Kulagin:2004ie,Kulagin:2010gd}.
The resulting prediction on $d/u$ ratio for the proton is similar to the one obtained in Ref.~\cite{Alekhin:2017kpj} without the use of any nuclear data.
The presence of nuclear $^2$H, $^3$He, and $^3$H DIS data in the QCD analysis allows a significant reduction of the uncertainty on the proton $d/u$ ratio at large $x$.
Further improvements are expected from \mara{} data on the ratios $\sigma^{\htri}/\sigma^d$ and $\sigma^{\hetri}/\sigma^d$~\cite{Petratos2019}.

We emphasize the importance of taking into account HT terms in the QCD analysis of DIS data with $Q^2 \lesssim 10$ GeV$^2$. 
Two different HT models are considered: additive (aHT) and multiplicative (mHT) HT models.
While the aHT model provides a good performance with isoscalar HT terms and the OS function, in the mHT model the HT terms are different for protons and neutrons, because of a correlation with the LT terms.
In the mHT model we also find a nonzero neutron-proton asymmetry in the OS function.
The ratio $d/u$ at large $x$ is enhanced in the mHT model as compared to that in the aHT model.
These results are driven by the \mara{} $\hetri/\htri$ data and originate from the interplay between the LT and HT terms in SFs, which is inherent to the mHT model.
We conclude that this feature of the mHT model can lead to potential biases and inconsistencies, while the \mara{} $\sigma^{\hetri}/\sigma^{\htri}$ data clearly prefer the aHT model over the mHT one with $\chi^2/\text{NDP}=20/22$ vs 34/22.

Future precision cross-section measurements with $^2$H, $^3$H and $^3$He targets in a wide kinematical region would allow to address the HT model and to further constrain the isospin dependence of nuclear effects at the parton level. More precise measurements of the latter will require future flavor sensitive data from DIS at the electron-ion collider~\cite{AbdulKhalek:2021gbh} and from both neutrino and antineutrino charged-current interactions off hydrogen and various isoscalar and non-isoscalar nuclear 
targets~\cite{Petti:2019asx,Petti:2022bzt,Petti:2023osk} at the long-baseline neutrino facility~\cite{DUNE:2020ypp}. 

We thank M.~V.~Garzelli and S.-O.~Moch for valuable comments,
G.~Salm\`e for providing the $\htri$ and $\hetri$ spectral functions of Ref.~\cite{Pace:2001cm},
and G.~G.~Petratos for clarifications about the \mara{} data.
S.~A. is supported by the DFG Grants No. MO 1801/5-1 and No. KN 365/14-1.
R.~P. thanks the support of USC and of the CERN neutrino platform.

\bibliography{references}

\begin{thebibliography}{31}%
\makeatletter
\providecommand \@ifxundefined [1]{%
 \@ifx{#1\undefined}
}%
\providecommand \@ifnum [1]{%
 \ifnum #1\expandafter \@firstoftwo
 \else \expandafter \@secondoftwo
 \fi
}%
\providecommand \@ifx [1]{%
 \ifx #1\expandafter \@firstoftwo
 \else \expandafter \@secondoftwo
 \fi
}%
\providecommand \natexlab [1]{#1}%
\providecommand \enquote  [1]{``#1''}%
\providecommand \bibnamefont  [1]{#1}%
\providecommand \bibfnamefont [1]{#1}%
\providecommand \citenamefont [1]{#1}%
\providecommand \href@noop [0]{\@secondoftwo}%
\providecommand \href [0]{\begingroup \@sanitize@url \@href}%
\providecommand \@href[1]{\@@startlink{#1}\@@href}%
\providecommand \@@href[1]{\endgroup#1\@@endlink}%
\providecommand \@sanitize@url [0]{\catcode `\\12\catcode `\$12\catcode `\&12\catcode `\#12\catcode `\^12\catcode `\_12\catcode `\%12\relax}%
\providecommand \@@startlink[1]{}%
\providecommand \@@endlink[0]{}%
\providecommand \url  [0]{\begingroup\@sanitize@url \@url }%
\providecommand \@url [1]{\endgroup\@href {#1}{\urlprefix }}%
\providecommand \urlprefix  [0]{URL }%
\providecommand \Eprint [0]{\href }%
\providecommand \doibase [0]{https://doi.org/}%
\providecommand \selectlanguage [0]{\@gobble}%
\providecommand \bibinfo  [0]{\@secondoftwo}%
\providecommand \bibfield  [0]{\@secondoftwo}%
\providecommand \translation [1]{[#1]}%
\providecommand \BibitemOpen [0]{}%
\providecommand \bibitemStop [0]{}%
\providecommand \bibitemNoStop [0]{.\EOS\space}%
\providecommand \EOS [0]{\spacefactor3000\relax}%
\providecommand \BibitemShut  [1]{\csname bibitem#1\endcsname}%
\let\auto@bib@innerbib\@empty
\bibitem [{\citenamefont {Alekhin}\ \emph {et~al.}(2022)\citenamefont {Alekhin}, \citenamefont {Kulagin},\ and\ \citenamefont {Petti}}]{Alekhin:2022tip}%
  \BibitemOpen
  \bibfield  {author} {\bibinfo {author} {\bibfnamefont {S.~I.}\ \bibnamefont {Alekhin}}, \bibinfo {author} {\bibfnamefont {S.~A.}\ \bibnamefont {Kulagin}},\ and\ \bibinfo {author} {\bibfnamefont {R.}~\bibnamefont {Petti}},\ }\href {https://doi.org/10.1103/PhysRevD.105.114037} {\bibfield  {journal} {\bibinfo  {journal} {Phys. Rev. D}\ }\textbf {\bibinfo {volume} {105}},\ \bibinfo {pages} {114037} (\bibinfo {year} {2022})},\ \Eprint {https://arxiv.org/abs/2203.07333} {arXiv:2203.07333 [hep-ph]} \BibitemShut {NoStop}%
\bibitem [{\citenamefont {Alekhin}\ \emph {et~al.}(2023)\citenamefont {Alekhin}, \citenamefont {Kulagin},\ and\ \citenamefont {Petti}}]{Alekhin:2022uwc}%
  \BibitemOpen
  \bibfield  {author} {\bibinfo {author} {\bibfnamefont {S.~I.}\ \bibnamefont {Alekhin}}, \bibinfo {author} {\bibfnamefont {S.~A.}\ \bibnamefont {Kulagin}},\ and\ \bibinfo {author} {\bibfnamefont {R.}~\bibnamefont {Petti}},\ }\href {https://doi.org/10.1103/PhysRevD.107.L051506} {\bibfield  {journal} {\bibinfo  {journal} {Phys. Rev. D}\ }\textbf {\bibinfo {volume} {107}},\ \bibinfo {pages} {L051506} (\bibinfo {year} {2023})},\ \Eprint {https://arxiv.org/abs/2211.09514} {arXiv:2211.09514 [hep-ph]} \BibitemShut {NoStop}%
\bibitem [{\citenamefont {Abrams}\ \emph {et~al.}(2022)\citenamefont {Abrams} \emph {et~al.}}]{MARATHON:2021vqu}%
  \BibitemOpen
  \bibfield  {author} {\bibinfo {author} {\bibfnamefont {D.}~\bibnamefont {Abrams}} \emph {et~al.},\ }\href {https://doi.org/10.1103/PhysRevLett.128.132003} {\bibfield  {journal} {\bibinfo  {journal} {Phys. Rev. Lett.}\ }\textbf {\bibinfo {volume} {128}},\ \bibinfo {pages} {132003} (\bibinfo {year} {2022})},\ \Eprint {https://arxiv.org/abs/2104.05850} {arXiv:2104.05850 [hep-ex]} \BibitemShut {NoStop}%
\bibitem [{\citenamefont {Alekhin}\ \emph {et~al.}(2017{\natexlab{a}})\citenamefont {Alekhin}, \citenamefont {Bl\"umlein}, \citenamefont {Moch},\ and\ \citenamefont {Placakyte}}]{Alekhin:2017kpj}%
  \BibitemOpen
  \bibfield  {author} {\bibinfo {author} {\bibfnamefont {S.}~\bibnamefont {Alekhin}}, \bibinfo {author} {\bibfnamefont {J.}~\bibnamefont {Bl\"umlein}}, \bibinfo {author} {\bibfnamefont {S.}~\bibnamefont {Moch}},\ and\ \bibinfo {author} {\bibfnamefont {R.}~\bibnamefont {Placakyte}},\ }\href {https://doi.org/10.1103/PhysRevD.96.014011} {\bibfield  {journal} {\bibinfo  {journal} {Phys. Rev. D}\ }\textbf {\bibinfo {volume} {96}},\ \bibinfo {pages} {014011} (\bibinfo {year} {2017}{\natexlab{a}})},\ \Eprint {https://arxiv.org/abs/1701.05838} {arXiv:1701.05838 [hep-ph]} \BibitemShut {NoStop}%
\bibitem [{\citenamefont {Kulagin}\ and\ \citenamefont {Petti}(2006)}]{Kulagin:2004ie}%
  \BibitemOpen
  \bibfield  {author} {\bibinfo {author} {\bibfnamefont {S.~A.}\ \bibnamefont {Kulagin}}\ and\ \bibinfo {author} {\bibfnamefont {R.}~\bibnamefont {Petti}},\ }\href {https://doi.org/10.1016/j.nuclphysa.2005.10.011} {\bibfield  {journal} {\bibinfo  {journal} {Nucl. Phys. A}\ }\textbf {\bibinfo {volume} {765}},\ \bibinfo {pages} {126} (\bibinfo {year} {2006})},\ \Eprint {https://arxiv.org/abs/hep-ph/0412425} {arXiv:hep-ph/0412425} \BibitemShut {NoStop}%
\bibitem [{\citenamefont {Akulinichev}\ \emph {et~al.}(1985)\citenamefont {Akulinichev}, \citenamefont {Kulagin},\ and\ \citenamefont {Vagradov}}]{Akulinichev:1985ij}%
  \BibitemOpen
  \bibfield  {author} {\bibinfo {author} {\bibfnamefont {S.~V.}\ \bibnamefont {Akulinichev}}, \bibinfo {author} {\bibfnamefont {S.~A.}\ \bibnamefont {Kulagin}},\ and\ \bibinfo {author} {\bibfnamefont {G.~M.}\ \bibnamefont {Vagradov}},\ }\href {https://doi.org/10.1016/0370-2693(85)90799-3} {\bibfield  {journal} {\bibinfo  {journal} {Phys. Lett. B}\ }\textbf {\bibinfo {volume} {158}},\ \bibinfo {pages} {485} (\bibinfo {year} {1985})}\BibitemShut {NoStop}%
\bibitem [{\citenamefont {Kulagin}(1989)}]{Kulagin:1989mu}%
  \BibitemOpen
  \bibfield  {author} {\bibinfo {author} {\bibfnamefont {S.~A.}\ \bibnamefont {Kulagin}},\ }\href {https://doi.org/10.1016/0375-9474(89)90233-9} {\bibfield  {journal} {\bibinfo  {journal} {Nucl. Phys. A}\ }\textbf {\bibinfo {volume} {500}},\ \bibinfo {pages} {653} (\bibinfo {year} {1989})}\BibitemShut {NoStop}%
\bibitem [{\citenamefont {Kulagin}\ \emph {et~al.}(1994)\citenamefont {Kulagin}, \citenamefont {Piller},\ and\ \citenamefont {Weise}}]{Kulagin:1994fz}%
  \BibitemOpen
  \bibfield  {author} {\bibinfo {author} {\bibfnamefont {S.~A.}\ \bibnamefont {Kulagin}}, \bibinfo {author} {\bibfnamefont {G.}~\bibnamefont {Piller}},\ and\ \bibinfo {author} {\bibfnamefont {W.}~\bibnamefont {Weise}},\ }\href {https://doi.org/10.1103/PhysRevC.50.1154} {\bibfield  {journal} {\bibinfo  {journal} {Phys. Rev. C}\ }\textbf {\bibinfo {volume} {50}},\ \bibinfo {pages} {1154} (\bibinfo {year} {1994})},\ \Eprint {https://arxiv.org/abs/nucl-th/9402015} {arXiv:nucl-th/9402015} \BibitemShut {NoStop}%
\bibitem [{\citenamefont {Alekhin}\ \emph {et~al.}(2017{\natexlab{b}})\citenamefont {Alekhin}, \citenamefont {Kulagin},\ and\ \citenamefont {Petti}}]{Alekhin:2017fpf}%
  \BibitemOpen
  \bibfield  {author} {\bibinfo {author} {\bibfnamefont {S.~I.}\ \bibnamefont {Alekhin}}, \bibinfo {author} {\bibfnamefont {S.~A.}\ \bibnamefont {Kulagin}},\ and\ \bibinfo {author} {\bibfnamefont {R.}~\bibnamefont {Petti}},\ }\href {https://doi.org/10.1103/PhysRevD.96.054005} {\bibfield  {journal} {\bibinfo  {journal} {Phys. Rev. D}\ }\textbf {\bibinfo {volume} {96}},\ \bibinfo {pages} {054005} (\bibinfo {year} {2017}{\natexlab{b}})},\ \Eprint {https://arxiv.org/abs/1704.00204} {arXiv:1704.00204 [nucl-th]} \BibitemShut {NoStop}%
\bibitem [{\citenamefont {Accardi}\ \emph {et~al.}(2016)\citenamefont {Accardi} \emph {et~al.}}]{Accardi:2016ndt}%
  \BibitemOpen
  \bibfield  {author} {\bibinfo {author} {\bibfnamefont {A.}~\bibnamefont {Accardi}} \emph {et~al.},\ }\href {https://doi.org/10.1140/epjc/s10052-016-4285-4} {\bibfield  {journal} {\bibinfo  {journal} {Eur. Phys. J. C}\ }\textbf {\bibinfo {volume} {76}},\ \bibinfo {pages} {471} (\bibinfo {year} {2016})},\ \Eprint {https://arxiv.org/abs/1603.08906} {arXiv:1603.08906 [hep-ph]} \BibitemShut {NoStop}%
\bibitem [{\citenamefont {Georgi}\ and\ \citenamefont {Politzer}(1976)}]{Georgi:1976ve}%
  \BibitemOpen
  \bibfield  {author} {\bibinfo {author} {\bibfnamefont {H.}~\bibnamefont {Georgi}}\ and\ \bibinfo {author} {\bibfnamefont {H.~D.}\ \bibnamefont {Politzer}},\ }\href {https://doi.org/10.1103/PhysRevD.14.1829} {\bibfield  {journal} {\bibinfo  {journal} {Phys. Rev. D}\ }\textbf {\bibinfo {volume} {14}},\ \bibinfo {pages} {1829} (\bibinfo {year} {1976})}\BibitemShut {NoStop}%
\bibitem [{\citenamefont {Virchaux}\ and\ \citenamefont {Milsztajn}(1992)}]{Virchaux:1991jc}%
  \BibitemOpen
  \bibfield  {author} {\bibinfo {author} {\bibfnamefont {M.}~\bibnamefont {Virchaux}}\ and\ \bibinfo {author} {\bibfnamefont {A.}~\bibnamefont {Milsztajn}},\ }\href {https://doi.org/10.1016/0370-2693(92)90527-B} {\bibfield  {journal} {\bibinfo  {journal} {Phys. Lett. B}\ }\textbf {\bibinfo {volume} {274}},\ \bibinfo {pages} {221} (\bibinfo {year} {1992})}\BibitemShut {NoStop}%
\bibitem [{\citenamefont {Kulagin}\ and\ \citenamefont {Petti}(2010)}]{Kulagin:2010gd}%
  \BibitemOpen
  \bibfield  {author} {\bibinfo {author} {\bibfnamefont {S.~A.}\ \bibnamefont {Kulagin}}\ and\ \bibinfo {author} {\bibfnamefont {R.}~\bibnamefont {Petti}},\ }\href {https://doi.org/10.1103/PhysRevC.82.054614} {\bibfield  {journal} {\bibinfo  {journal} {Phys. Rev. C}\ }\textbf {\bibinfo {volume} {82}},\ \bibinfo {pages} {054614} (\bibinfo {year} {2010})},\ \Eprint {https://arxiv.org/abs/1004.3062} {arXiv:1004.3062 [hep-ph]} \BibitemShut {NoStop}%
\bibitem [{\citenamefont {Wiringa}\ \emph {et~al.}(1995)\citenamefont {Wiringa}, \citenamefont {Stoks},\ and\ \citenamefont {Schiavilla}}]{Wiringa:1994wb}%
  \BibitemOpen
  \bibfield  {author} {\bibinfo {author} {\bibfnamefont {R.~B.}\ \bibnamefont {Wiringa}}, \bibinfo {author} {\bibfnamefont {V.~G.~J.}\ \bibnamefont {Stoks}},\ and\ \bibinfo {author} {\bibfnamefont {R.}~\bibnamefont {Schiavilla}},\ }\href {https://doi.org/10.1103/PhysRevC.51.38} {\bibfield  {journal} {\bibinfo  {journal} {Phys. Rev. C}\ }\textbf {\bibinfo {volume} {51}},\ \bibinfo {pages} {38} (\bibinfo {year} {1995})},\ \Eprint {https://arxiv.org/abs/nucl-th/9408016} {arXiv:nucl-th/9408016 [nucl-th]} \BibitemShut {NoStop}%
\bibitem [{\citenamefont {Veerasamy}\ and\ \citenamefont {Polyzou}(2011)}]{Veerasamy:2011ak}%
  \BibitemOpen
  \bibfield  {author} {\bibinfo {author} {\bibfnamefont {S.}~\bibnamefont {Veerasamy}}\ and\ \bibinfo {author} {\bibfnamefont {W.~N.}\ \bibnamefont {Polyzou}},\ }\href {https://doi.org/10.1103/PhysRevC.84.034003} {\bibfield  {journal} {\bibinfo  {journal} {Phys. Rev. C}\ }\textbf {\bibinfo {volume} {84}},\ \bibinfo {pages} {034003} (\bibinfo {year} {2011})},\ \Eprint {https://arxiv.org/abs/1106.1934} {arXiv:1106.1934 [nucl-th]} \BibitemShut {NoStop}%
\bibitem [{\citenamefont {Pace}\ \emph {et~al.}(2001)\citenamefont {Pace}, \citenamefont {Salme}, \citenamefont {Scopetta},\ and\ \citenamefont {Kievsky}}]{Pace:2001cm}%
  \BibitemOpen
  \bibfield  {author} {\bibinfo {author} {\bibfnamefont {E.}~\bibnamefont {Pace}}, \bibinfo {author} {\bibfnamefont {G.}~\bibnamefont {Salme}}, \bibinfo {author} {\bibfnamefont {S.}~\bibnamefont {Scopetta}},\ and\ \bibinfo {author} {\bibfnamefont {A.}~\bibnamefont {Kievsky}},\ }\href {https://doi.org/10.1103/PhysRevC.64.055203} {\bibfield  {journal} {\bibinfo  {journal} {Phys. Rev. C}\ }\textbf {\bibinfo {volume} {64}},\ \bibinfo {pages} {055203} (\bibinfo {year} {2001})},\ \Eprint {https://arxiv.org/abs/nucl-th/0109005} {arXiv:nucl-th/0109005} \BibitemShut {NoStop}%
\bibitem [{\citenamefont {Alekhin}\ \emph {et~al.}(2007)\citenamefont {Alekhin}, \citenamefont {Kulagin},\ and\ \citenamefont {Petti}}]{Alekhin:2007fh}%
  \BibitemOpen
  \bibfield  {author} {\bibinfo {author} {\bibfnamefont {S.}~\bibnamefont {Alekhin}}, \bibinfo {author} {\bibfnamefont {S.~A.}\ \bibnamefont {Kulagin}},\ and\ \bibinfo {author} {\bibfnamefont {R.}~\bibnamefont {Petti}},\ }\href {https://doi.org/10.1063/1.2834481} {\bibfield  {journal} {\bibinfo  {journal} {AIP Conf. Proc.}\ }\textbf {\bibinfo {volume} {967}},\ \bibinfo {pages} {215} (\bibinfo {year} {2007})},\ \Eprint {https://arxiv.org/abs/0710.0124} {arXiv:0710.0124 [hep-ph]} \BibitemShut {NoStop}%
\bibitem [{\citenamefont {Aaij}\ \emph {et~al.}(2015{\natexlab{a}})\citenamefont {Aaij} \emph {et~al.}}]{LHCb:2015kwa}%
  \BibitemOpen
  \bibfield  {author} {\bibinfo {author} {\bibfnamefont {R.}~\bibnamefont {Aaij}} \emph {et~al.} (\bibinfo {collaboration} {LHCb}),\ }\href {https://doi.org/10.1007/JHEP05(2015)109} {\bibfield  {journal} {\bibinfo  {journal} {J. High Energy Phys.}\ }\textbf {\bibinfo {volume} {2015}}\bibfield  {number} {\bibinfo  {number} { (05)},\ \bibinfo {pages} {109}},\ }\Eprint {https://arxiv.org/abs/1503.00963} {arXiv:1503.00963 [hep-ex]} \BibitemShut {NoStop}%
\bibitem [{\citenamefont {Aaij}\ \emph {et~al.}(2015{\natexlab{b}})\citenamefont {Aaij} \emph {et~al.}}]{LHCb:2015okr}%
  \BibitemOpen
  \bibfield  {author} {\bibinfo {author} {\bibfnamefont {R.}~\bibnamefont {Aaij}} \emph {et~al.} (\bibinfo {collaboration} {LHCb}),\ }\href {https://doi.org/10.1007/JHEP08(2015)039} {\bibfield  {journal} {\bibinfo  {journal} {J. High Energy Phys.}\ }\textbf {\bibinfo {volume} {2015}}\bibfield  {number} {\bibinfo  {number} { (08)},\ \bibinfo {pages} {039}},\ }\Eprint {https://arxiv.org/abs/1505.07024} {arXiv:1505.07024 [hep-ex]} \BibitemShut {NoStop}%
\bibitem [{\citenamefont {Aaij}\ \emph {et~al.}(2016)\citenamefont {Aaij} \emph {et~al.}}]{LHCb:2015mad}%
  \BibitemOpen
  \bibfield  {author} {\bibinfo {author} {\bibfnamefont {R.}~\bibnamefont {Aaij}} \emph {et~al.} (\bibinfo {collaboration} {LHCb}),\ }\href {https://doi.org/10.1007/JHEP01(2016)155} {\bibfield  {journal} {\bibinfo  {journal} {J. High Energy Phys.}\ }\textbf {\bibinfo {volume} {2016}}\bibfield  {number} {\bibinfo  {number} { (01)},\ \bibinfo {pages} {155}},\ }\Eprint {https://arxiv.org/abs/1511.08039} {arXiv:1511.08039 [hep-ex]} \BibitemShut {NoStop}%
\bibitem [{\citenamefont {Abazov}\ \emph {et~al.}(2015)\citenamefont {Abazov} \emph {et~al.}}]{D0:2014kma}%
  \BibitemOpen
  \bibfield  {author} {\bibinfo {author} {\bibfnamefont {V.~M.}\ \bibnamefont {Abazov}} \emph {et~al.} (\bibinfo {collaboration} {D0}),\ }\href {https://doi.org/10.1103/PhysRevD.91.032007} {\bibfield  {journal} {\bibinfo  {journal} {Phys. Rev. D}\ }\textbf {\bibinfo {volume} {91}},\ \bibinfo {pages} {032007} (\bibinfo {year} {2015})},\ \bibinfo {note} {[Erratum: Phys. Rev. D 91, 079901 (2015)]},\ \Eprint {https://arxiv.org/abs/1412.2862} {arXiv:1412.2862 [hep-ex]} \BibitemShut {NoStop}%
\bibitem [{\citenamefont {Kulagin}\ and\ \citenamefont {Petti}(2014)}]{Kulagin:2014vsa}%
  \BibitemOpen
  \bibfield  {author} {\bibinfo {author} {\bibfnamefont {S.~A.}\ \bibnamefont {Kulagin}}\ and\ \bibinfo {author} {\bibfnamefont {R.}~\bibnamefont {Petti}},\ }\href {https://doi.org/10.1103/PhysRevC.90.045204} {\bibfield  {journal} {\bibinfo  {journal} {Phys. Rev. C}\ }\textbf {\bibinfo {volume} {90}},\ \bibinfo {pages} {045204} (\bibinfo {year} {2014})},\ \Eprint {https://arxiv.org/abs/1405.2529} {arXiv:1405.2529 [hep-ph]} \BibitemShut {NoStop}%
\bibitem [{\citenamefont {Kulagin}(2017)}]{Kulagin:2016fzf}%
  \BibitemOpen
  \bibfield  {author} {\bibinfo {author} {\bibfnamefont {S.~A.}\ \bibnamefont {Kulagin}},\ }\href {https://doi.org/10.1051/epjconf/201713801006} {\bibfield  {journal} {\bibinfo  {journal} {EPJ Web Conf.}\ }\textbf {\bibinfo {volume} {138}},\ \bibinfo {pages} {01006} (\bibinfo {year} {2017})},\ \Eprint {https://arxiv.org/abs/1612.07741} {arXiv:1612.07741 [hep-ph]} \BibitemShut {NoStop}%
\bibitem [{\citenamefont {Cocuzza}\ \emph {et~al.}(2021)\citenamefont {Cocuzza}, \citenamefont {Keppel}, \citenamefont {Liu}, \citenamefont {Melnitchouk}, \citenamefont {Metz}, \citenamefont {Sato},\ and\ \citenamefont {Thomas}}]{Cocuzza:2021rfn}%
  \BibitemOpen
  \bibfield  {author} {\bibinfo {author} {\bibfnamefont {C.}~\bibnamefont {Cocuzza}}, \bibinfo {author} {\bibfnamefont {C.~E.}\ \bibnamefont {Keppel}}, \bibinfo {author} {\bibfnamefont {H.}~\bibnamefont {Liu}}, \bibinfo {author} {\bibfnamefont {W.}~\bibnamefont {Melnitchouk}}, \bibinfo {author} {\bibfnamefont {A.}~\bibnamefont {Metz}}, \bibinfo {author} {\bibfnamefont {N.}~\bibnamefont {Sato}},\ and\ \bibinfo {author} {\bibfnamefont {A.~W.}\ \bibnamefont {Thomas}},\ }\href {https://doi.org/10.1103/PhysRevLett.127.242001} {\bibfield  {journal} {\bibinfo  {journal} {Phys. Rev. Lett.}\ }\textbf {\bibinfo {volume} {127}},\ \bibinfo {pages} {242001} (\bibinfo {year} {2021})},\ \Eprint {https://arxiv.org/abs/2104.06946} {arXiv:2104.06946 [hep-ph]} \BibitemShut {NoStop}%
\bibitem [{\citenamefont {Segarra}\ \emph {et~al.}(2021)\citenamefont {Segarra} \emph {et~al.}}]{Segarra:2021exb}%
  \BibitemOpen
  \bibfield  {author} {\bibinfo {author} {\bibfnamefont {E.~P.}\ \bibnamefont {Segarra}} \emph {et~al.},\ }\Eprint {https://arxiv.org/abs/2104.07130} {arXiv:2104.07130 [hep-ph]}  (\bibinfo {year} {2021})\BibitemShut {NoStop}%
\bibitem [{\citenamefont {Petratos}(2019)}]{Petratos2019}%
  \BibitemOpen
  \bibfield  {author} {\bibinfo {author} {\bibfnamefont {G.~G.}\ \bibnamefont {Petratos}},\ }\href@noop {} {}\bibinfo {howpublished} {\href{https://indico.cern.ch/event/799284/}{talk at the HiX2019 Workshop}} (\bibinfo {year} {2019})\BibitemShut {NoStop}%
\bibitem [{\citenamefont {Abdul~Khalek}\ \emph {et~al.}(2022)\citenamefont {Abdul~Khalek} \emph {et~al.}}]{AbdulKhalek:2021gbh}%
  \BibitemOpen
  \bibfield  {author} {\bibinfo {author} {\bibfnamefont {R.}~\bibnamefont {Abdul~Khalek}} \emph {et~al.},\ }\href {https://doi.org/10.1016/j.nuclphysa.2022.122447} {\bibfield  {journal} {\bibinfo  {journal} {Nucl. Phys. A}\ }\textbf {\bibinfo {volume} {1026}},\ \bibinfo {pages} {122447} (\bibinfo {year} {2022})},\ \Eprint {https://arxiv.org/abs/2103.05419} {arXiv:2103.05419 [physics.ins-det]} \BibitemShut {NoStop}%
\bibitem [{\citenamefont {Petti}(2019)}]{Petti:2019asx}%
  \BibitemOpen
  \bibfield  {author} {\bibinfo {author} {\bibfnamefont {R.}~\bibnamefont {Petti}},\ }\Eprint {https://arxiv.org/abs/1910.05995} {arXiv:1910.05995 [hep-ex]}  (\bibinfo {year} {2019})\BibitemShut {NoStop}%
\bibitem [{\citenamefont {Petti}(2022)}]{Petti:2022bzt}%
  \BibitemOpen
  \bibfield  {author} {\bibinfo {author} {\bibfnamefont {R.}~\bibnamefont {Petti}},\ }\href {https://doi.org/10.1016/j.physletb.2022.137469} {\bibfield  {journal} {\bibinfo  {journal} {Phys. Lett. B}\ }\textbf {\bibinfo {volume} {834}},\ \bibinfo {pages} {137469} (\bibinfo {year} {2022})},\ \Eprint {https://arxiv.org/abs/2205.10396} {arXiv:2205.10396 [hep-ph]} \BibitemShut {NoStop}%
\bibitem [{\citenamefont {Petti}(2023)}]{Petti:2023osk}%
  \BibitemOpen
  \bibfield  {author} {\bibinfo {author} {\bibfnamefont {R.}~\bibnamefont {Petti}},\ }\href@noop {} {\  (\bibinfo {year} {2023})},\ \Eprint {https://arxiv.org/abs/2301.04744} {arXiv:2301.04744 [hep-ex]} \BibitemShut {NoStop}%
\bibitem [{\citenamefont {Abi}\ \emph {et~al.}(2020)\citenamefont {Abi} \emph {et~al.}}]{DUNE:2020ypp}%
  \BibitemOpen
  \bibfield  {author} {\bibinfo {author} {\bibfnamefont {B.}~\bibnamefont {Abi}} \emph {et~al.} (\bibinfo {collaboration} {DUNE}),\ }\Eprint {https://arxiv.org/abs/2002.03005} {arXiv:2002.03005 [hep-ex]}  (\bibinfo {year} {2020})\BibitemShut {NoStop}%
\end{thebibliography}%
\end{document}